\documentclass[fleqn,usenatbib]{mnras} 

\usepackage{newtxtext,newtxmath}

\usepackage[T1]{fontenc}
\usepackage{ae,aecompl}

\usepackage{graphicx}	
\usepackage{amsmath}	
\usepackage{amssymb}	
\usepackage{times}

\usepackage{commath}
\usepackage{tikz,pgfplots}
\usepackage{upgreek}
\usepackage{natbib}

\usepackage{hyperref}

\usetikzlibrary{pgfplots.groupplots}

\defcitealias{Zhao1996}{Z96}
\defcitealias{LSE}{LSEE}

\newcommand{\expe}{\mathrm{e}}
\newcommand{\Beta}{\mathcal{B}}

\DeclareMathOperator\cosec{cosec}

\title[Double-power-law biorthonormal expansions]{A two-parameter family of double-power-law biorthonormal potential-density expansions}

\author[E. Lilley et al.]{
Edward J. Lilley,$^{1}$\thanks{E-mail: ejl44,jls,nwe@cam.ac.uk}
Jason L. Sanders,$^{1}$
N. Wyn Evans$^{1,2}$
\\
$^{1}$Institute of Astronomy, Madingley Rd, Cambridge, CB3 0HA\\
$^{2}$Center for Computational Astrophysics, Flatiron Institute, 162 5th Avenue, New York, NY10010, USA
}

\date{Accepted XXX. Received YYY; in original form ZZZ}

\pubyear{2017}

\begin{document}
\label{firstpage}
\pagerange{\pageref{firstpage}--\pageref{lastpage}}
\maketitle

\begin{abstract}
We present a two-parameter family of biorthonormal double-power-law potential-density expansions. Both the potential and density are given in closed analytic form and may be rapidly computed via recurrence relations.
We show that this family encompasses all the known analytic biorthonormal expansions: the Zhao expansions (themselves generalizations of ones found earlier by Hernquist \& Ostriker and by Clutton-Brock) and the recently discovered \citet{LSE} expansion. Our new two-parameter family includes expansions based around many familiar spherical density profiles as zeroth-order models, including the $\gamma$ models and the Jaffe model. It also contains a basis expansion that reproduces the famous Navarro-Frenk-White (NFW) profile at zeroth order. The new basis expansions have been found via a systematic methodology which has wide applications in finding other new expansions. In the process, we also uncovered a novel integral transform solution to Poisson's equation.
\end{abstract}

\begin{keywords}
galaxies: haloes -- galaxies: structure
-- methods: numerical
\end{keywords}


\section{Introduction}

There has been renewed interest in basis function or halo expansion techniques in recent years. Historically, basis functions were introduced to study problems in galactic stability~\citep{Fr84} or to provide numerical algorithms to evolve collisionless stellar systems~\citep{hernquist1992}. An influential paper by \citet{lowing2011} suggested a brand new application, namely that the technique can provide an efficient description of the structure of numerical dark matter haloes, as well as their evolution. This opens up the possibility of repeatedly re-running the costly original simulation using the basis functions to study the fate of tidal streams or small satellite galaxies~\citep[e.g.,][]{Ngan2015}.

The existing basis function expansions have been found piecemeal and seemingly by inspired guesswork. First, \citet{cluttonbrock1973} and then \citet{hernquist1992} identified biorthogonal expansions whose lowest order model is the \citet{Plummer1911} or \citet{Hernquist1990} sphere respectively. Subsequently, \citet[hereafter Z96]{Zhao1996} found a neat way of incorporating them into a one-parameter sequence whose lowest order models are the hypervirial family~\citep{Evans2005}. More generally,
\citet{Wein99} pointed out that an expansion with lowest order basis function for any spherical model can be computed by numerical solution of the Sturm-Liouville equation. Very recently, \citet[hereafter LSEE] {LSE} identified a completely new set of analytic biorthogonal expansions based on a lowest order model with density $\rho \sim r^{1/\alpha-2}$ at small radii and $\rho \sim r^{-3-1/(2\alpha)}$ at large radii ($\alpha\geq1/2$). For $\alpha=1$, this provides a close analogue to the well-known \citet*[NFW]{NFW1997} profile of cold dark matter haloes \citep{Lilley2017b} with the sobriquet 'the super-NFW model'. \citetalias{LSE}'s expansion also incorporates an earlier, isolated result of \cite{rahmati2009} on setting $\alpha=1/2$. There are some striking similarities between the two biorthogonal expansions (\citetalias{Zhao1996} and \citetalias{LSE}) that strongly suggest that they are part of an underlying and more complete theoretical framework.  It is the purpose of this paper to provide it.

All of the known spherical basis expansions have double power law density profiles at lowest order. A general analytic double power law model for the density profile of galaxies is
\begin{equation}
\rho(r)\propto r^{-\gamma}(1+r^{1/\alpha})^{-(\beta-\gamma)\alpha},
\label{eq:doublepowlaw}
\end{equation}
where the three parameters $(\alpha,\beta,\gamma)$ describe the turn-over, outer slope and inner slope and we have chosen units such that the scale-length is unity. The corresponding potentials are simple, reducing to elementary functions for the four cases discussed by \cite{Zhao1996}, and they present a widely-used way to model the gravitational field of a galaxy. Deviations away from this smooth model are efficiently captured using a series of biorthogonal potential-density pairs. These pairs of functions $(\Phi_{(nlm)},\rho_{(nlm)})$ are indexed by the integer tuple $(n,l,m)$ (where $n\geq0$, $l\geq0$ and $|m|\leq l$) and satisfy
\begin{equation}
\int\mathrm{d}^3\boldsymbol{r}\,\Phi_{nlm}(\boldsymbol{r})\rho_{n'l'm'}(\boldsymbol{r})=4\pi N_{nl}\delta_{nn'}\delta_{ll'}\delta_{mm'},
\end{equation}
for some choice of normalization $N_{nl}$.
The angular parts of the basis functions are expanded in terms of the spherical harmonics (normalized to have $4\pi$ weight)
\begin{equation}
\begin{split}
&\Phi_{nlm}(r,\theta,\phi)=\Phi_{nl}(r)Y_{lm}(\theta,\phi),\\
&\rho_{nlm}(r,\theta,\phi)=\rho_{nl}(r)Y_{lm}(\theta,\phi),
\end{split}
\label{eq:sharmonics}
\end{equation}
such that the potential-density pair $(\Phi_{nl},\rho_{nl})$ satisfy the Poisson equation
\begin{equation}
\Big(\nabla^2-\frac{l(l+1)}{r^2}\Big)\Phi_{nl}(r)=4\pi G K_{nl}\rho_{nl}(r),
\label{eqn::poisson}
\end{equation}
given some constant $K_{nl}$ and the orthogonality relation
\begin{equation}
\int\mathrm{d}r\,r^2\Phi_{nl}(r)\rho_{n'l}(r)=N_{nl}\delta_{nn'}.
\label{eqn::orthogonal}
\end{equation}
From now on, we set $G=1$. These expansions have been used extensively to efficiently model the shapes of galaxies away from smooth spherical models as well as in $N$-body models to reduce two-body effects in the computation of the force. 

The \citetalias{Zhao1996} and \citetalias{LSE} basis function expansions both have double-power-law density profiles at lowest order. The two families of models are quite distinct and lie along completely separate curves in the 3d space spanned by $(\alpha,\beta,\gamma)$. The \citetalias{Zhao1996} sequence is defined by   $\beta=3+1/\alpha$, $\gamma=2-1/\alpha$, whilst the \citetalias{LSE} sequence lies along $\beta=3+1/(2\alpha)$, $\gamma=2-1/\alpha$. It is therefore natural to ask whether these two basis expansions can be encompassed as special cases of a more general family of biorthogonal potential-density expansions that covers more of the $(\alpha,\beta,\gamma)$ space.
 
Here, we present a two-parameter family of expansions that encompasses all the known closed form biorthogonal potential-density pairs. Section~\ref{Section::NonOrtho} demonstrates how to construct a non-orthonormal basis through the Hankel transform which reproduces double-power-law density profiles at lowest order. The non-orthonormal set is diagonalized analytically producing an orthonormal set in Section~\ref{Section::Ortho}. Special cases, including the cosmologically significant NFW model, are discussed in Section~\ref{Section::Special}. This paper deals with the theoretical framework, but we provide elsewhere an efficient numerical implementation for the NFW model, together with applications.

\section{A non-orthonormal basis set}\label{Section::NonOrtho}

\subsection{Family A}

Following \citet{LSE}, we begin by writing a solution for the potential and density basis functions in equation~\eqref{eqn::poisson} as
\begin{equation}
\begin{split}
&\Phi_{nl}(r) \propto r^{-1/2} \int_0^\infty \dif k \: g_n(k) \: J_{\mu}(kz), \\
&\rho_{nl}(r) \propto r^{1/\alpha - 5/2} \int_0^\infty \dif k \: k^2 \: g_n(k) \: J_{\mu}(kz),
\end{split}
\label{eqn::familyA}
\end{equation}
where $z=r^{1/(2\alpha)}$ and $\mu=\alpha(1+2l)$. We refer to this set of solutions as \emph{Family A} and will present a further family in the next subsection. The orthogonality condition of equation~\eqref{eqn::orthogonal} is only satisfied if 
\begin{equation}
\int_0^\infty\dif k k g_m(k) g_n(k) 
\propto\delta_{mn}.
\end{equation}
Given a density basis function $\rho_{nl}(r)$, $g_n(k)$ is found by inverting the Hankel transform as
\begin{equation}
g_n(k) = k^{-1}\int_0^\infty \dif z \: z \: r^{5/2-1/\alpha} \rho_{nl}(r) \: J_{\mu}(kz).
\end{equation}
For instance, using the zeroth order \citetalias{Zhao1996} basis function, 
\begin{equation}
\rho_{0l}(r)\propto r^{-5/2+1/\alpha}\frac{z^\mu}{(1+z^2)^{\mu+2}},
\end{equation}
the inversion gives \citep[][6.565(4)]{gradshteyn4}
\begin{equation}
g_0(k)=k^\mu K_{1}(k),
\end{equation}
where $K_\nu(k)$ is the modified Bessel function of the second kind \citep[][(10.25), satisfying the identity $K_{-\nu}(k)=K_\nu(k)$]{dlmf}. This leads us to propose a generalized form for $g_0(k)$ as
\begin{equation}
g_0(k)=k^{\mu+\nu-1}  K_{\nu}(k),
\label{eq:gGen}
\end{equation}
which produces the zeroth order density functions of
\begin{equation}
\rho_{0l}(r)\propto\frac{r^{1/\alpha + l - 2}}{(1+r^{1/\alpha})^{\mu+\nu+1}}.
\label{eqn::zerothdensity}
\end{equation}
using \citet[][6.576(7)]{gradshteyn4} and potential functions of 
\begin{equation}
\Phi_{0l}(r) \propto r^{-1/2}z^\mu{}_2 \mathcal{F}_1\left(\mu,\mu+\nu;1+\mu;-z^2\right)
\propto \frac{\Beta_\chi(\mu,\nu)}{r^{l+1}}
\label{eqn::zerothpotential}
\end{equation}
Here, $\chi=z^2/(1+z^2)$, $\Beta_x(a,b)$ is the incomplete beta function and we have used the integral 6.576(3) from \citet{gradshteyn4} and the linear hypergeometric transformation \citep[][15.8.1]{dlmf}. The potential integral is only valid for $\mu+\nu>0$, but this constraint is less restrictive than the orthogonality constraint on $\mu$ and $\nu$ (discussed in the following section). The potential basis functions recover the required $r^l$ behaviour for $r\rightarrow0$ and $r^{-1-l}$ for $r\rightarrow\infty$ ~\citep[see e.g.,][]{hernquist1992,LSE}. The inner density slope is $\gamma=2-1/\alpha$ whilst the outer density slope is $\beta=3+\nu/\alpha$. For a $\gamma=1$ cusp, $\alpha=1$ and $\nu$ controls the outer slope. Slower breaks (e.g. $\alpha=2$) produce cuspier ($\gamma>1$) central profiles. To avoid unphysical centrally-vanishing density profiles we require $\alpha\geq1/2$ and in turn if we require profiles with finite mass ($\beta>3$) then $\nu>0$.

In the top panel of Figure~\ref{fig::example_profiles}, we show the range of zeroth-order density profiles encompassed by our Family A of models. We see increasing $\alpha$ at fixed $\nu$ `straightens out' the density profile whilst increasing $\nu$ at fixed $\alpha$ steepens the outer density slope.

We now wish to construct a full basis set with this lowest order potential-density pair. Computing $g_1(k)$ from the first order density basis function of the \citetalias{Zhao1996} basis set gives
\begin{equation}
g_1(k)=k^\mu (k K_0(k)-\mu K_{1}(k)),
\end{equation} 
suggesting that a full set of solutions can be composed from the set of non-orthonormal basis functions
\begin{equation}\label{eqn::curly_K}
\mathcal{K}_j(k)=k^{\mu+\nu-1+j}K_{\nu-j}(k), \qquad j>0, \:\: j\in\mathbb{Z}.
\end{equation}
The corresponding non-biorthonormal potential-density basis functions $(\tilde\Phi_{nl},\tilde\rho_{nl})$ can be found by applying \citet[][6.576(3)]{gradshteyn4},
%
%
\begin{equation}
\begin{split}
{\tilde\Phi}_{nl} 
&\propto \frac{r^l}{(1+z^2)^{\mu+\nu}}\mathcal{P}^{(\nu)}_{j-1}\!(\chi),\\
\tilde\rho_{nl}
&\propto\frac{r^{l+1/\alpha-2}}{(1+z^2)^{\mu+\nu+1}}\mathcal{P}^{(\nu+1)}_j\!(\chi),
\end{split}
\label{equation::nonortho}
\end{equation}
where we use the shorthand $\mathcal{P}^{(\nu)}_j\!(\chi)$ for a certain hypergeometric polynomial which can be computed directly as a Jacobi polynomial
%
%
\begin{multline}\label{eqn::curly_P}
\mathcal{P}^{(\nu)}_j\!(\chi) \equiv {}_2 \mathcal{F}_1\left(-j,\mu+\nu;1+\mu;\chi\right) = \frac{(-1)^j j!}{(\mu+1)_j} P^{(\nu-1-j,\mu)}_j\left(\xi\right), \\
\xi \equiv 2\chi-1 = \frac{z^2-1}{z^2+1}.
\end{multline}
and we have made use of the Pochhammer symbol $(z)_n$ \citep{Ab72}.
The only term in the expressions \eqref{equation::nonortho} which is
not proportional to a polynomial in $\chi$ is the zeroth-order ($j=0$) of the
potential, given by equation \eqref{eqn::zerothpotential} in terms of the incomplete beta function.

\subsection{Family B}

A further solution to the Poisson equation, similar to equation~\eqref{eqn::familyA}, is
\begin{equation}
\begin{split}
&\Phi_{nl}(r) \propto r^{-1/2} \int_0^\infty \dif k \: g_n(k) \: J_{\mu}(k/z),\\
&\rho_{nl}(r) \propto r^{-1/\alpha - 5/2} \int_0^\infty \dif k \: k^2 \: g_n(k) \: J_{\mu}(k/z),
\end{split}
\end{equation}
where the difference to equation~\eqref{eqn::familyA} is in the argument of the Bessel functions. With the same choice of $g_n(k)$ as in equation~\eqref{eq:gGen}, we find
\begin{equation}
\rho_{0l}(r) \propto \frac{r^{\nu/\alpha + l - 2}}{(1 + r^{1/\alpha})^{\mu + \nu + 1}},
\end{equation}
\begin{equation}
\Phi_{0l}(r) \propto r^l \: \Beta_{1 - \chi}(\mu, \nu).
\end{equation}
The inner density slope is $\gamma=2-\nu/\alpha$ whilst the outer density slope is $\beta=3+1/\alpha$. For cusped models ($0 < \nu < 2 \alpha$), $\alpha$ controls the outer slope but also alters the turn-over of the density profile. We call this family of models \emph{Family B}. The potential integral is only valid for $\mu+\nu>0$. For non-vanishing central density, we require $\nu<2\alpha$. All zeroth-order models have finite mass as $\alpha>0$. Note that for $\nu=1$, Family A and Family B coincide and provide the \citetalias{Zhao1996} solutions, special cases of which include the \citet{cluttonbrock1973} and \citet{hernquist1992} expansions.  However, in general, Family B is distinct from Family A, even if the models have the same inner $\gamma$ and outer $\beta$ density slopes. This is because the gradualness of the transition from inner to outer behaviour is controlled by $\alpha$, which is in general different between the two families.

\begin{figure}
$$\includegraphics[width=0.5\textwidth]{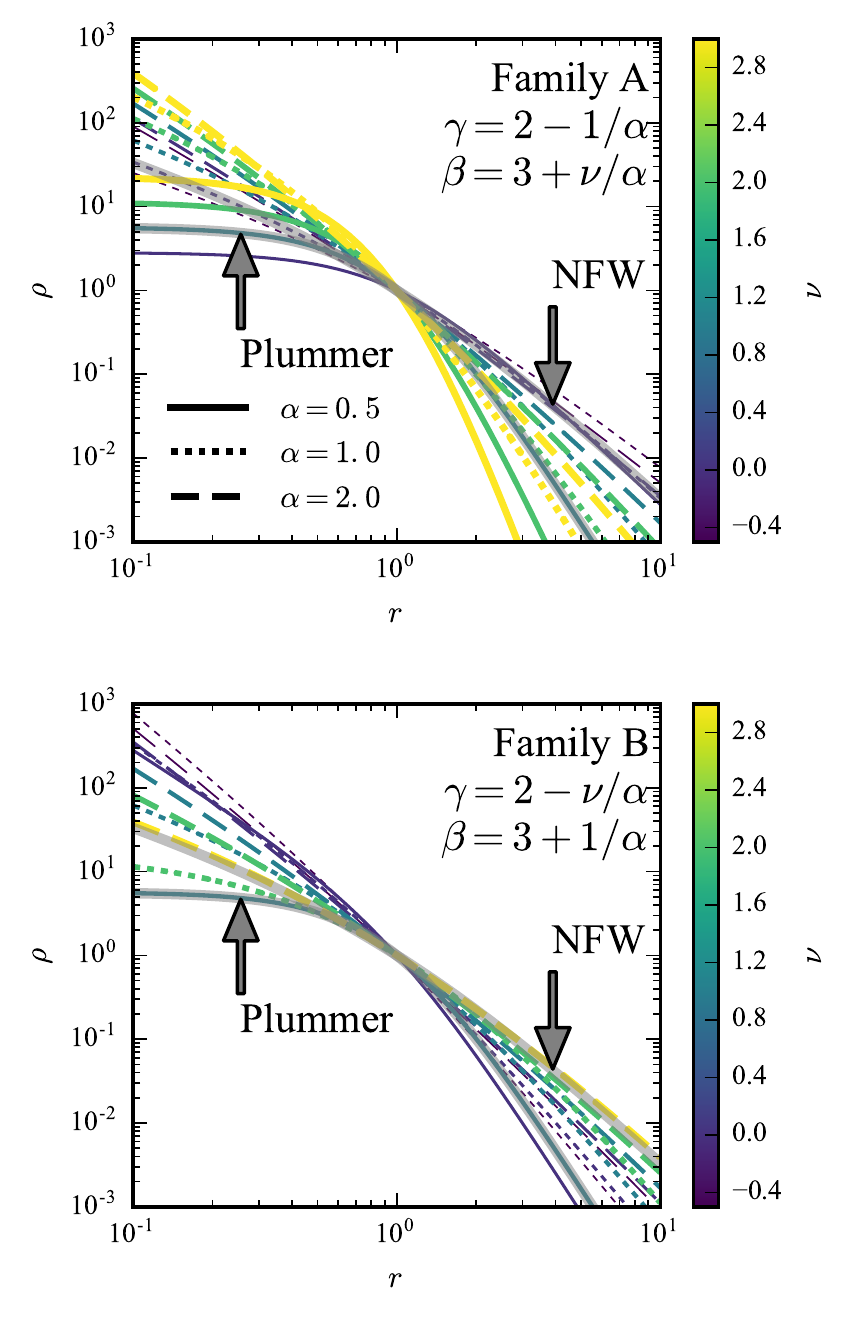}$$
\caption{The range of zeroth-order density profiles covered by our two families of expansions (A top, B bottom). Each line is coloured by the value of $\nu$ and the line-styles give the $\alpha$ values. In light grey, we show a Plummer profile and NFW profile.}
\label{fig::example_profiles}
\end{figure}

In the bottom panel of Figure~\ref{fig::example_profiles}, we show the range of zeroth-order density profiles in Family B. We see that increasing $\alpha$ at fixed $\nu$ `straightens out' the density profile as with Family B, whilst increasing $\nu$ at fixed $\alpha$ steepens the inner density profile.

\section{An orthonormal basis set}\label{Section::Ortho}

To find an orthonormal basis set, we construct a linear sum of the non-orthonormal basis as
\begin{equation}
g_n(k)=\sum_{j=0}^{n}c_{nj}\:\mathcal{K}_j(k),
\label{eqn::gk_with_K}
\end{equation}
subject to the orthonormality requirement
\begin{equation}\label{eqn:gnk_orthog}
\int_0^\infty\dif k k g_m(k) g_n(k) 
=\delta_{mn}.
\end{equation}
To evaluate $c_{nj}$, we require the integral \citep[][6.576(4)]{gradshteyn4}, indicating by $\Beta(a,b)$ the (complete) beta function, 
\begin{multline}\label{eqn::Koverlap}
D_{mn}(\mu,\nu)\equiv\int_0^\infty\dif k k \mathcal{K}_m(k)\mathcal{K}_n(k) \\
= 2^{m + n + 2\mu + 2\nu - 3}\Gamma(m + \mu + \nu) \Gamma(n + \mu + \nu)\Beta(m+n+\mu,\mu + 2 \nu).
\end{multline}
We note that this integral only converges when $\mu>-2\nu$ as each potential-density inner product is required to be finite. To see this directly for the zeroth-order case, the following integral must be finite,
\begin{equation}
\int_0^\infty \dif r \: r^2 \: \Phi_{00} \: \rho_{00} \propto \int_0^\infty \dif r \: \frac{\Beta_\chi(\alpha,\nu)}{r} \: \frac{r^{1/\alpha}}{(1 + r^{1/\alpha})^{\alpha + \nu + 1}}.
\label{eqn::zeroth_order_inner_product}
\end{equation}
As $r \to \infty$, we have $\chi \approx 1 - r^{-1/\alpha}$, so we can approximate the incomplete beta function's defining integral as
\begin{equation}
\Beta_\chi(\alpha, \nu) \approx \Beta(\mu,\nu) - r^{-\nu/\alpha}.
\end{equation}
Hence the asymptotic behaviour of the zeroth order potential function is
\begin{equation}
\Phi_{00} \sim
\begin{cases}
  r^{-1}, & \mathrm{if} \: \nu/\alpha \geq 0 \\
  r^{-\nu/\alpha - 1}, & \mathrm{otherwise}.
\end{cases}
\end{equation}
Inspecting the behaviour of the integrand in Eq.~\ref{eqn::zeroth_order_inner_product} as $r \to \infty$ for
Family A ($\alpha \geq 1/2$) we find that if $\nu \geq 0$ then the
integral clearly converges. However, if $\nu < 0$ then to prevent divergence we
must have $\alpha > -2\nu$. An identical constraint on $\alpha$ and $\nu$ is obtained for Family B by considering $r \to 0$.


Although it may appear that a numerical inversion of the matrix
\eqref{eqn::Koverlap} must be performed, a closed-form expression can
in fact be found. Taking advantage of the beta function's integral representation,
\begin{equation}\label{eqn::beta_repn}
\Beta(m+n+\mu,\mu + 2 \nu) = \int_0^1\mathrm{d}t\,t^{m+n+\mu-1}(1-t)^{\mu+2\nu-1},
\end{equation}
and replacing $\mathcal{K}_n$ in \eqref{eqn::Koverlap} by some linear
combination $\sum c_{jn} \mathcal{K}_n$, we see that the orthogonality
condition \eqref{eqn:gnk_orthog} becomes an orthogonality relation
between two polynomials in $t$, with respect to a certain weight
function,
\begin{equation}
\int_0^1\mathrm{d}t\,t^{\mu-1}(1-t)^{\mu+2\nu-1}\Bigg(\sum_{m=0}^{i}c_{im}t^m\Bigg)\Bigg(\sum_{n=0}^{j}c_{jn}t^n\Bigg) \propto \delta_{ij}.
\end{equation}
 Fortunately the orthogonal polynomials corresponding to this weight function are well-known: they are simply the Jacobi polynomials combined with a linear change of variables, namely $P_n^{(\mu+2\nu-1,\mu-1)}\left(2t-1\right)$. A simple closed-form expression for these polynomials as a sum over monomials in $t$ can be obtained via the representation found in \citet[8.962(1)]{gradshteyn4},
\begin{equation}\label{eqn::jacobi}
  P^{(\mu+2\nu-1,\mu-1)}_n\left(2t-1\right) = \frac{(-1)^n \: (\mu)_n}{n!} \sum_{j=0}^n \frac{(-n)_j \: (n+2\mu+2\nu-1)_j}{j! \: (\mu)_j} \: t^j.
\end{equation}
Writing the quantities $c_{jn}$ in terms of the coefficients of this
polynomial gives us an expression for $g_n(k)$ in integral form;
inserting this expression into \eqref{eqn::familyA} and then using an
approach based on generating functions (detailed in Appendix
\ref{sec::general_derivation}) gives a simple recurrence relation for
the potential basis functions and a closed form for the density basis
functions. 
Recalling the shorthands $\mu = \alpha
(1+2l)$, $z = r^{1/(2\alpha)}$, $\chi = z^2/(1+z^2)$ and $\xi =
2\chi-1$, we have for the potential,
\begin{equation}\label{eqn::phi_hat}
\begin{split}
{\Phi}_{nl} - {\Phi}_{n+1,l} & = \frac{2 \: n!}{(\mu+1)_n} \: \frac{r^l}{(1+z^2)^{\mu+\nu}} \: P_n^{(\mu+2\nu-1,\mu)}(\xi), \\
{\Phi}_{0l} & = \frac{\mu \: \Beta_\chi(\mu,\nu)}{r^{1+l}}, \\
\end{split}
\end{equation}
and for the density,
\begin{equation}\label{eqn::rho_hat}
\begin{split}
    {\rho}_{nl} &= \frac{r^{l-2+1/\alpha}}{(1+z^2)^{\mu+\nu+1}}%
  \left[a_{nl}P_n^{(\mu+2\nu-1,\mu)}(\xi)
    - b_{nl}P_{n-1}^{(\mu+2\nu-1,\mu)}(\xi)\right], \\
  a_{nl} &= (n+2\mu+2\nu-1)(n+\mu+\nu), \\
  b_{nl} &= (n+\mu+2\nu-1)(n+\mu+\nu-1).
\end{split}
\end{equation}
The normalisation constant (which is derived from the normalisation of the Jacobi polynomials \eqref{eqn::jacobi}) is
\begin{equation}
N_{nl} = \frac{\alpha\Gamma(n+\mu+2\nu)\Gamma(\mu+1)}{\Gamma(n+2\mu+2\nu-1)},
\end{equation}
and the proportionality constant in Poisson's equation is
\begin{equation}
K_{nl} = -\frac{n!\Gamma(\mu+1)}{4\pi\alpha^2(2n+2\mu+2\nu-1)\Gamma(n+\mu)}.
\end{equation}
Note that limiting forms of the basis functions and associated constants must be used for the case $\alpha + \nu = 1/2$, for which see Appendix \ref{sec::limiting}.

The basis functions of Family B can be constructed from those of Family A by the transformations: $\chi\rightarrow1-\chi$, $\xi\rightarrow-\xi$, 
${\rho}_{nl}\rightarrow{\rho}_{nl}r^{(\nu-1)/\alpha}$, ${\Phi}_{0l}\rightarrow{\Phi}_{0l}r^{1+2l}$ and $({\Phi}_{nl}-{\Phi}_{0l})\rightarrow({\Phi}_{nl}-{\Phi}_{0l})r^{\nu/\alpha}$. 
We again emphasise that Families A and B are in general distinct, other than for the ($\nu=1$) sequence of models given in \citetalias{Zhao1996}.

The family of basis sets described by Eqns \eqref{eqn::phi_hat} and \eqref{eqn::rho_hat} (and the accompanying `B' sets) are the major result of this paper. By choosing $\alpha$ and $\nu$ appropriately, they can be used to efficiently capture the higher-order corrections to a double-power-law model with any combination of inner and outer slopes. The basis sets are analytical -- they require no further numerical orthogonalisation, and hence the resulting accuracy is not dependent on the condition number of an overlap matrix (compare \citet{Saha1993}, where this orthogonalisation step must be carried out).

\begin{figure*}
\begin{minipage}{0.45\textwidth}
$$\includegraphics[width=\textwidth]{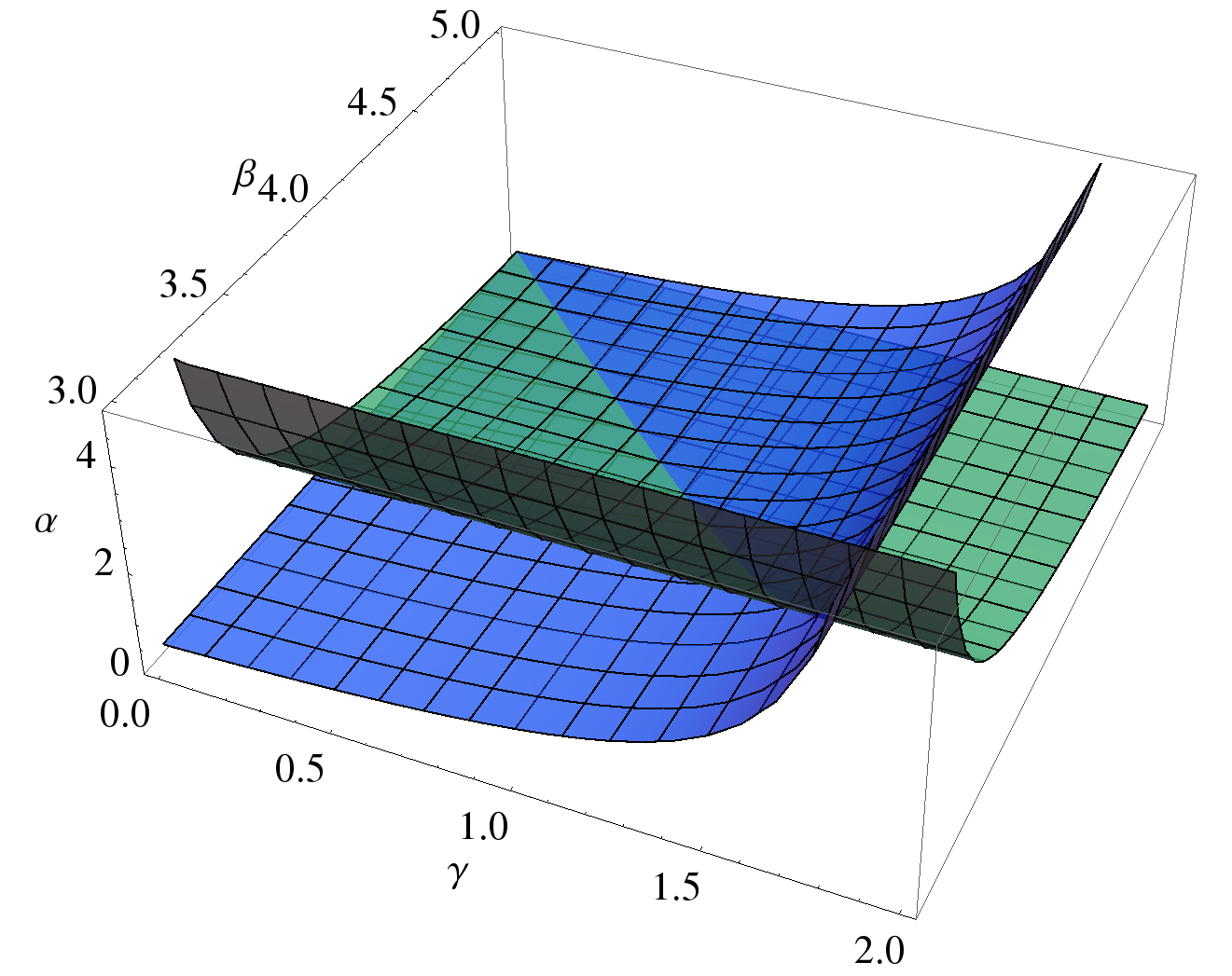}$$
\end{minipage}
\begin{minipage}{0.45\textwidth}
$$\includegraphics[width=\textwidth]{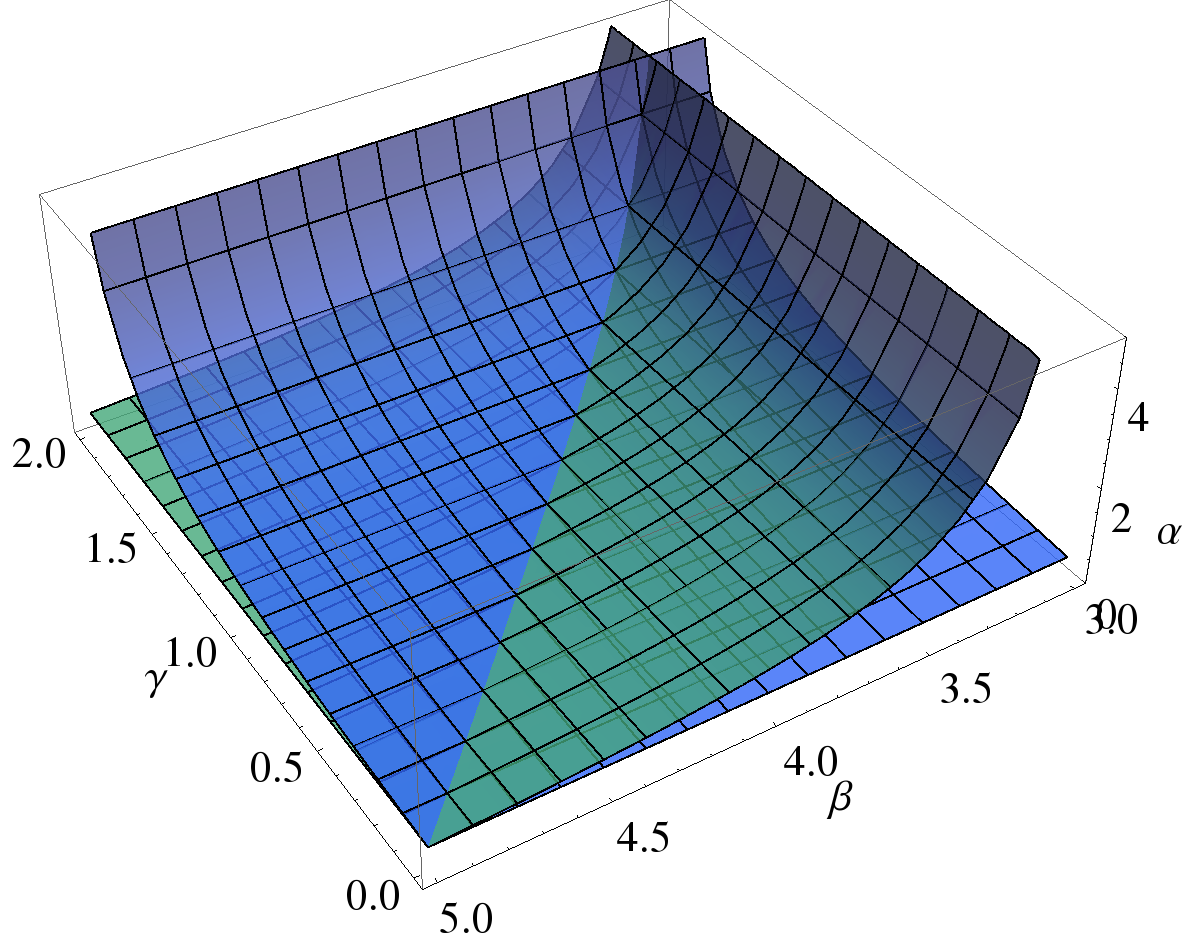}$$
\end{minipage}
$$\includegraphics[width=0.8\textwidth]{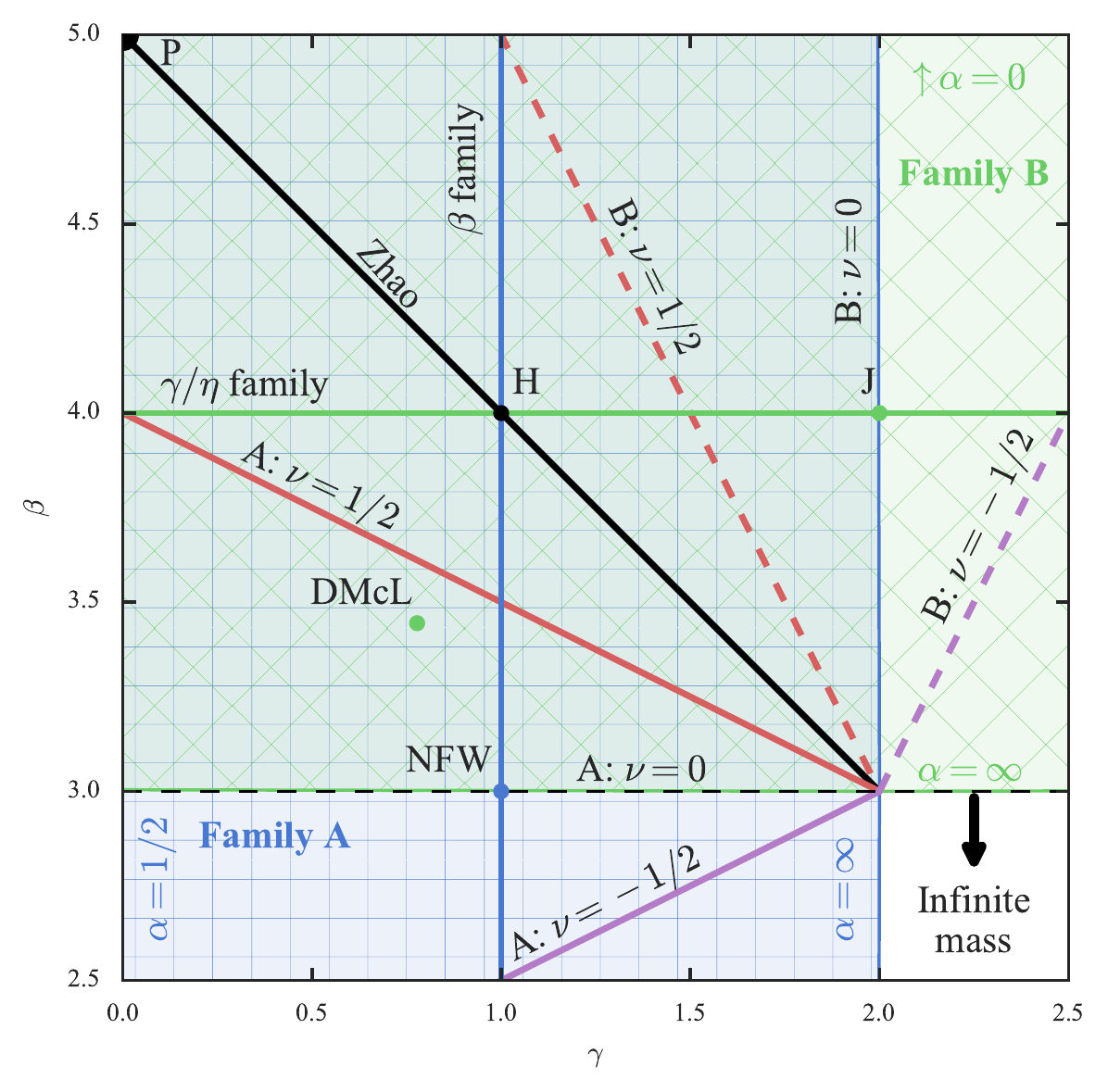}$$
\caption{Upper Panels: Plots of the surfaces of Family A (blue) and Family B (green) in ($\alpha,\beta,\gamma$) parameter space. The intersection of the two surfaces is the \citetalias{Zhao1996} sequence. Lower Panel: Range of inner $\gamma$ and outer $\beta$ slopes encompassed by our basis expansion (blue square shading for Family A and green diagonal shading for Family B). This is the projection of the surfaces in the upper panels into the ($\beta,\gamma$) plane. Subsets of these families are marked with solid lines: black shows the \citetalias{Zhao1996} sequence (Family A and B), red and purple shows the \citetalias{LSE} sequence (Family A, $\nu = \pm 1/2$). The red and purple dashed lines show the sequence on Family B with $\nu = \pm 1/2$. The blue vertical line shows Zhao's $\beta$ sequence in Family A, whilst the green horizontal line the $\gamma$ models of \citet{Dehnen1993} and \citet{Tremaine1994} in Family B. Five specific models are shown by points: the NFW, the Plummer (P), the Hernquist (H), the Jaffe (J) and the Dehnen and McLaughlin (DMcL). The colour of the point indicates the Family in which they reside. For all these models, the methods of this paper allow us to construct biorthogonal basis function expansions.}
\label{fig::solution_space}
\end{figure*}

\section{Special cases}\label{Section::Special}

Our two-parameter family of expansions encompasses a number of well-known zeroth-order models as well as all the previously known families of biorthogonal 3D basis expansions as special cases. In Figure~\ref{fig::solution_space}, we show the range of inner and outer slopes accessible with our two families of models along with the known families and other well-known zeroth-order models. We will discuss each of these known limits before presenting the new special cases encompassed by our family. Each special case is obtained from our general expressions \eqref{eqn::phi_hat} and \eqref{eqn::rho_hat} by setting the appropriate value of $\nu$.

\subsection{The Zhao (Z96) Sequence ($\nu=1$)}

\citetalias{Zhao1996} gives a family of basis sets whose zeroth orders correspond to his `$\alpha$'-family of simple analytical potential-density pairs (also known as \citet{Ve79} or hypervirial \citep{Evans2005} models). This sequence of basis sets fits into our scheme by setting $\nu=1$ and letting $\alpha$ remain arbitrary in either Family A or Family B.

In this case, both the density and potential basis functions reduce to Gegenbauer polynomials multiplied by the zeroth order term in the expansion,
\begin{equation}
\Phi_{nl}(r)\propto r^{-1/2}\frac{z^\mu}{(1+z^2)^{\mu}}C^{(\mu+1/2)}_n(\xi),
\end{equation}
\begin{equation}
\rho_{nl}(r)\propto r^{-5/2+1/\alpha}\frac{z^\mu}{(1+z^2)^{\mu+2}}C^{(\mu+1/2)}_n(\xi),
\end{equation}
This covers the Plummer profile ($\alpha=1/2$) and Hernquist profile ($\alpha=1$), as first derived by \citet{cluttonbrock1973} and \citet{hernquist1992} respectively.

\subsection{The LSEE Sequence ($\nu=\pm1/2$)}

When $\nu= \pm 1/2$ in Family A, we recover the \citetalias{LSE}
expansion. Using the properties of modified Bessel functions of
half-integer order \citep[10.47.9,10.49.16]{dlmf}, we see that for
$\nu=\pm1/2$, $\mathcal{K}_0(k)$ is proportional to
$k^{\mu}\mathrm{e}^{-k}$. Up to a factor of $k$ this is the weight
function for the associated Laguerre polynomials, so natural choices
for $g_n(k)$ (see \eqref{eqn::familyA}) are
\begin{equation}
g_n(k) = k^{\mu-1} \: \expe^{-k} \: L_n^{(2\mu-1)}(2k), \qquad \nu=1/2,
\end{equation}
\begin{equation*}
g_n(k) = k^{\mu-2} \: \expe^{-k} \: L_n^{(2\mu-3)}(2k), \qquad \nu=-1/2.
\end{equation*}

%
%

When $\nu=\alpha=1/2$ the zeroth order is the perfect sphere of
\citet{dezeeuw1985}, as first derived by \citet{rahmati2009}. When
$\nu=1/2$ and $\alpha =1$, the zeroth order is the super-Hernquist model, which has
a cosmologically important $1/r$ density cusp at the centre
\citep{Lilley2017b}.

\subsection{NFW and associated models ($\nu=0$)}\label{sec::nfw}

When we set $\nu = 0$, we obtain Family A expansions whose
lowest-order densities all have outer slope $\beta = 3$, and Family B expansions
with inner slope  $\gamma = 2$. This set encompasses a number of well-studied and astrophysically interesting profiles.
%
For example, when $\alpha = 1$ the beta function in the family A
potential can be expressed as a logarithm, revealing the well-known
NFW potential and density \citep{NFW1997}
\begin{equation}
\rho_{00} \propto \frac{1}{r(1+r)^2},\qquad\qquad \Phi_{00} \propto \frac{-\ln{(1+r)}}{r}.
\end{equation}
Furthermore, setting $\alpha=1/2$ for Family A we produce a basis set whose zeroth order is the modified Hubble profile and setting $\alpha=1$ for Family B we find the zeroth order model is the \citet{Ja83} profile.

See Section \ref{sec::beta_functions} for a note on computing the zeroth-order potential for this family of basis sets.



\subsection{Elementary subsets of the double power-law family}

\citetalias{Zhao1996} shows that there are four cases when the potentials of the double power-law family (\ref{eq:doublepowlaw}) reduce to simpler analytic functions. These occur when combinations of $(\alpha, \beta, \gamma)$ take on integer values (we will use $k$ and $k'$ as integers).

The `$\alpha$' subset is obtained when $(\alpha,\beta,\gamma)=(\alpha,3+k'/\alpha,2-k/\alpha)$ with the `$\alpha$' family corresponding to $k=k'=1$. Family A contains the members of the `$\alpha$' subset with $k=1$ by choosing integer $\nu$ and Family B contains the members with $k'=1$ also by choosing integer $\nu$. A related subset is obtained when $(\alpha,\beta,\gamma)=(\alpha,2+k'/\alpha,3-k/\alpha)$. Family A contains the members of this subset with $k'=\alpha+\nu$ and $k=1+\alpha$ restricting both $\alpha$ and $\nu$ to integer values. Similarly, Family B contains the members with $k'=1+\alpha$ and $k=\alpha+\nu$.

A further elementary subset is the `$\gamma$' subset where $(\alpha,\beta,\gamma)=(k,3+k'/k,\gamma)$. This subset contains the special case of the so-called $\gamma$ models \citep{Dehnen1993,Tremaine1994} when $k=k'=1$. Our Family B encompasses the set of models with $k'=1$ by setting $\alpha=k$ and leaving $\nu$ arbitrary. The final elementary subset is denoted the `$\beta$' subset by
\citetalias{Zhao1996} where $(\alpha,\beta,\gamma)=(k',\beta,2-k/k')$. Family A encompasses the set of models with $k=1$ by setting $\alpha=k'$ and leaving $\nu$ arbitrary. The special case of the `$\beta$' family when $k'=k=1$ is discussed in more detail by \citealt{Zhao1996}.

Although \citetalias{Zhao1996} identifies these further subsets of the double-power-law family as possessing elementary potentials, he does not provide the corresponding biorthonormal basis sets. These are now accessible through our work.

Finally, we note that choosing $\alpha=9/4$ and $\nu=11/4$ for Family B we reproduce the \citet{De05} models at zeroth order.

\section{Numerical implementation}

\subsection{Beta functions}\label{sec::beta_functions}

In order to evaluate the zeroth-order potential \eqref{eqn::zerothpotential} numerically we need a numerical implementation of the incomplete beta function $\Beta_\chi(\mu,\nu)$ that covers the full parameter space. Common implementations of the incomplete beta function (e.g. GSL) only cover the case of strictly positive parameters $\mu,\nu$; we have $\mu \geq 1/2$ always, but must deal with the cases of zero or negative $\nu$.

When $-1<\nu<0$, we can manipulate the incomplete beta function as
\begin{equation}
\Beta_\chi(p,q)=\Beta_\chi(p,q+1)\frac{\Beta(p,q)}{\Beta(p,q+1)}-\frac{\chi^p(1-\chi)^q}{q},
\end{equation}
and use
\begin{equation}
\Beta(p,q)=\frac{\Gamma(p)\Gamma(q+1)}{\Gamma(p+q+1)}\frac{p+q}{q}\text{ for }q<0.
\end{equation}
For $\nu=0$, we must use a numerical implementation of the hypergeometric function, using the identity
\begin{equation}
\Beta_\chi(\mu,0) = \frac{\chi^\mu}{\mu} {}_2 \mathcal{F}_1\left(\mu,1;\mu+1;\chi\right),
\end{equation}
or any equivalent transformation \citep[][8.17.7]{dlmf}, unless
$2\alpha$ is an integer (such as in the NFW case), in which case the
incomplete beta function reduces to elementary functions at $l=0$ and
the higher-$l$ functions can be found using a recurrence formula
\citep[][8.17.20]{dlmf}.


\subsection{Jacobi polynomials}
To evaluate the higher order potential and density basis functions, we require a numerical implementation of the Jacobi polynomials $P^{(a,b)}_n(x)$. Our basis expansions are only valid for $a,b>-1$ which coincides with the domain of applicability in many numerical implementations. It is efficient to use a recursion relation satisfied by the Jacobi polynomials to construct the ladder of basis functions \citet[8.961.2]{gradshteyn4}
\begin{equation}
\begin{split}
2(&n + 1)(n + a + b + 1)(2n + a + b)P^{(a,b)}_{n+1}(x)
= \\&(2n + a + b + 1) 
\Big[(2n + a + b)(2n + a + b + 2)x + a^2 − b^2\Big]
P^{(a,b)}_n(x)\\&
-2(n + a)(n + b)(2n + a + b + 2)P^{(a,b)}_{n-1}(x),
\end{split}
\end{equation}
with the lowest order polynomials given by
\begin{equation}
P^{(a,b)}_0(x)=1; 
\:\:\:
P^{(a,b)}_1(x)=\frac{1}{2}(a-b+(2+a+b)x).
\end{equation}
For the forces, we require the derivatives of Jacobi polynomials which are simply given by \citet[8.961.4]{gradshteyn4}
\begin{equation}
\frac{\mathrm{d}}{\mathrm{d}x}P^{(a,b)}_n(x) = \frac{1}{2}(n+a+b+1)P^{(a+1,b+1)}_{n-1}(x).
\end{equation}
Full computation of the forces requires recursive construction of two families of Jacobi polynomials $P^{(a,b)}_n(x)$ and $P^{(a+1,b+1)}_n(x)$.

\subsection{Numerical properties of the potential recurrence relation}
The ladder of potential basis functions for increasing $n$ is built up using the three-term inhomogeneous recurrence relation \eqref{eqn::phi_hat}. As $n \to \infty$ the terms in this relation tend to zero (and the rate at which this happens increases greatly with increasing $l$). This causes the computation of the potential functions to become inaccurate when $n$ is high (due to the accuracy with which the beta function in the zeroth-order basis function can be computed). We can remedy this using the same method as \citet{LSE} (see Section 4.1 of that paper for details). We pick some high order $N_\mathrm{max}$ for which the RHS of eq.~\eqref{eqn::phi_hat} is presumed to be approximately zero; then recurse backwards, constructing the ladder of Jacobi polynomials with decreasing $n$. This avoids the issue of cancellation of large terms.

\section{Conclusions}

The biorthonormal expansion series discovered by \citet{hernquist1992} has sometimes seemed miraculous. It has found widespread applications in astronomy \citep[e.g.,][]{Ba92,lowing2011,Ngan2015}. This is because the zeroth order potential-density pair is the spherical \citet{Hernquist1990} model, which is a reasonable representation of galaxies and dark haloes. The expansion enables us to describe deviations from sphericity (like triaxiality or lopsidedness) very easily. The biorthonormality ensures that the expansion coefficients for both the potential and the density can be calculated easily from an N-body realization.

This paper has studied the existence of biorthonormal basis function expansion methods for the general double-power-law family of density profiles. They are parameterised by $(\alpha,\beta,\gamma)$, where $\beta$ and $\gamma$ are the (negative) logarithmic gradients of the central and asymptotic profile, whilst $\alpha$ controls the briskness of the transition between inner and outer behaviour. We have presented an algorithm for constructing biorthonormal basis function expansions for two distinct families in $(\alpha,\beta,\gamma)$ space and provided closed analytic forms for the basis functions which may be efficiently computed via recursion relations. These results systematize all previously known biorthonormal basis function expansions for the spherical geometry, as discovered by \citet{cluttonbrock1973}, \citet{hernquist1992}, \citet{Zhao1996}, \citet{rahmati2009} and \citet{LSE}. It also provides new expansions for a host of familiar models, including the $\gamma$ models of \citet{Dehnen1993} and \citet{Tremaine1994}, the \citet{De05} models and the \citet{Ja83} model. Particularly significant in view of its cosmological importance is the \citet*{NFW1997} or NFW model. 

The work employs a methodical search for new biorthonormal basis expansions, unlike the inspired guesswork inherent in previous approaches. It is likely that our methodology can be followed to construct biorthonormal expansions for still more general zeroth-order potential-density pairs. In addition to the Bessel function solutions to the spherical Poisson equation \eqref{eqn::familyA}, we have demonstrated that the spherical Poisson equation can be solved by a novel integral transform technique involving confluent hypergeometric functions (see Appendix \ref{sec::general_derivation}, in particular equations \eqref{eqn::phi_int_form} and \eqref{eqn::rho_int_form}).

Our families of expansions lie along surfaces in the three-dimensional $(\alpha,\beta,\gamma)$ space. It is natural to ask whether our approach can be extended to cover the full 3D volume. Here, we suggest how to proceed based on the methodology employed in this paper. If we modify the $t$-dependent part of the integrand in \eqref{eqn::rho_int_form} to read $t^{\mu-1+n} \: \expe^{-t} \:  {}_1 \mathcal{F}_1\left(\lambda+1,\mu+1+n;t\right)$, we find that the associated density basis functions are
\begin{equation}
\rho_{nl} \propto  \frac{r^{l-2-\lambda/\alpha}}{(1+z^2)^{\mu+\nu-\lambda}} \: %
{}_2 \mathcal{F}_1\left(\left.%
     \begin{matrix}
       -n,\mu+\nu-\lambda \\
       \mu-\lambda
     \end{matrix}%
   \right| \chi \right),
\end{equation}
which generalises the non-biorthonormal density functions \eqref{equation::nonortho} to a three-parameter non-biorthonormal family whose zeroth-order has the double-power law form \eqref{eq:doublepowlaw} with inner slope $\gamma=2+\lambda/\alpha$ and outer slope $\beta=3+\nu/\alpha$, and with higher-order terms that simply multiply the zeroth-order by a polynomial. However, the continuation of our previous method requires that the overlap integral $\int r^2 \mathrm{d}r\, \rho_{nl}(r) \Phi_{n'l}(r)$ be expressible in a form that can be easily diagonalised, and this may be challenging.
Nonetheless, it seems likely that -- in addition to our Families A and B -- further sequences exist for which the procedure can be analytically carried out.

Although we have concentrated on theoretical matters here,
our discovery of an explicit set of entirely analytic biorthonormal basis functions for the NFW model has many astrophysical applications. It enables the distortions of dark halos to be described as higher order terms around the zeroth order NFW model. Elsewhere we provide a sampler of reconstructions of N-body haloes, as well as computer code that implements our numerical algorithm for the basis functions.

\section*{Acknowledgments}
EL and JLS acknowledge the support of the STFC. NWE thanks the Centre for Computational Astrophysics for hospitality during a working visit. We also thank members of the Institute of Astronomy Streams group for discussions and comments as this work was in progress.


\bibliographystyle{mnras}
\bibliography{bibliography}

\begin{thebibliography}{}
\makeatletter
\relax
\def\mn@urlcharsother{\let\do\@makeother \do\$\do\&\do\#\do\^\do\_\do\%\do\~}
\def\mn@doi{\begingroup\mn@urlcharsother \@ifnextchar [ {\mn@doi@}
  {\mn@doi@[]}}
\def\mn@doi@[#1]#2{\def\@tempa{#1}\ifx\@tempa\@empty \href
  {http://dx.doi.org/#2} {doi:#2}\else \href {http://dx.doi.org/#2} {#1}\fi
  \endgroup}
\def\mn@eprint#1#2{\mn@eprint@#1:#2::\@nil}
\def\mn@eprint@arXiv#1{\href {http://arxiv.org/abs/#1} {{\tt arXiv:#1}}}
\def\mn@eprint@dblp#1{\href {http://dblp.uni-trier.de/rec/bibtex/#1.xml}
  {dblp:#1}}
\def\mn@eprint@#1:#2:#3:#4\@nil{\def\@tempa {#1}\def\@tempb {#2}\def\@tempc
  {#3}\ifx \@tempc \@empty \let \@tempc \@tempb \let \@tempb \@tempa \fi \ifx
  \@tempb \@empty \def\@tempb {arXiv}\fi \@ifundefined
  {mn@eprint@\@tempb}{\@tempb:\@tempc}{\expandafter \expandafter \csname
  mn@eprint@\@tempb\endcsname \expandafter{\@tempc}}}

\bibitem[\protect\citeauthoryear{{Abramowitz} \& {Stegun}}{{Abramowitz} \&
  {Stegun}}{1972}]{Ab72}
{Abramowitz} M.,  {Stegun} I.~A.,  1972, {Handbook of Mathematical Functions}

\bibitem[\protect\citeauthoryear{{Barnes} \& {Hernquist}}{{Barnes} \&
  {Hernquist}}{1992}]{Ba92}
{Barnes} J.~E.,  {Hernquist} L.,  1992, \mn@doi [\araa]
  {10.1146/annurev.aa.30.090192.003421}, \href
  {http://adsabs.harvard.edu/abs/1992ARA%26A..30..705B} {30, 705}

\bibitem[\protect\citeauthoryear{Chaundy}{Chaundy}{1943}]{Chaundy1943}
Chaundy T.~W.,  1943, \mn@doi [The Quarterly Journal of Mathematics]
  {10.1093/qmath/os-14.1.55}, os-14, 55

\bibitem[\protect\citeauthoryear{{Clutton-Brock}}{{Clutton-Brock}}{1973}]{cluttonbrock1973}
{Clutton-Brock} M.,  1973, \mn@doi [\apss] {10.1007/BF00647652}, \href
  {http://adsabs.harvard.edu/abs/1973Ap%26SS..23...55C} {23, 55}

\bibitem[\protect\citeauthoryear{{Dehnen}}{{Dehnen}}{1993}]{Dehnen1993}
{Dehnen} W.,  1993, \mn@doi [\mnras] {10.1093/mnras/265.1.250}, \href
  {http://adsabs.harvard.edu/abs/1993MNRAS.265..250D} {265, 250}

\bibitem[\protect\citeauthoryear{{Dehnen} \& {McLaughlin}}{{Dehnen} \&
  {McLaughlin}}{2005}]{De05}
{Dehnen} W.,  {McLaughlin} D.~E.,  2005, \mn@doi [\mnras]
  {10.1111/j.1365-2966.2005.09510.x}, \href
  {http://adsabs.harvard.edu/abs/2005MNRAS.363.1057D} {363, 1057}

\bibitem[\protect\citeauthoryear{{Evans} \& {An}}{{Evans} \&
  {An}}{2005}]{Evans2005}
{Evans} N.~W.,  {An} J.~H.,  2005, \mn@doi [\mnras]
  {10.1111/j.1365-2966.2005.09078.x}, \href
  {http://adsabs.harvard.edu/abs/2005MNRAS.360..492E} {360, 492}

\bibitem[\protect\citeauthoryear{{Fridman} \& {Polyachenko}}{{Fridman} \&
  {Polyachenko}}{1984}]{Fr84}
{Fridman} A.~M.,  {Polyachenko} 1984, {Physics of gravitating systems. I.
  Equilibrium and stability.}

\bibitem[\protect\citeauthoryear{{Gradshteyn} \& {Ryzhik}}{{Gradshteyn} \&
  {Ryzhik}}{2014}]{gradshteyn4}
{Gradshteyn} I.,  {Ryzhik} I.,  2014, {Table of Integrals, Series and Products
  (Eighth Edition)}.
Academic Press

\bibitem[\protect\citeauthoryear{{Hernquist}}{{Hernquist}}{1990}]{Hernquist1990}
{Hernquist} L.,  1990, \mn@doi [\apj] {10.1086/168845}, \href
  {http://adsabs.harvard.edu/abs/1990ApJ...356..359H} {356, 359}

\bibitem[\protect\citeauthoryear{{Hernquist} \& {Ostriker}}{{Hernquist} \&
  {Ostriker}}{1992}]{hernquist1992}
{Hernquist} L.,  {Ostriker} J.~P.,  1992, \mn@doi [\apj] {10.1086/171025},
  \href {http://adsabs.harvard.edu/abs/1992ApJ...386..375H} {386, 375}

\bibitem[\protect\citeauthoryear{{Jaffe}}{{Jaffe}}{1983}]{Ja83}
{Jaffe} W.,  1983, \mn@doi [\mnras] {10.1093/mnras/202.4.995}, \href
  {http://adsabs.harvard.edu/abs/1983MNRAS.202..995J} {202, 995}

\bibitem[\protect\citeauthoryear{{Lilley}, {Evans}  \& {Sanders}}{{Lilley}
  et~al.}{2018a}]{Lilley2017b}
{Lilley} E.~J.,  {Evans} N.~W.,   {Sanders} J.~L.,  2018a, \mn@doi [\mnras]
  {10.1093/mnras/sty295}, \href
  {http://adsabs.harvard.edu/abs/2018MNRAS.476.2086L} {476, 2086}

\bibitem[\protect\citeauthoryear{{Lilley}, {Sanders}, {Evans}  \&
  {Erkal}}{{Lilley} et~al.}{2018b}]{LSE}
{Lilley} E.~J.,  {Sanders} J.~L.,  {Evans} N.~W.,   {Erkal} D.,  2018b, \mn@doi
  [\mnras] {10.1093/mnras/sty296}, \href
  {http://adsabs.harvard.edu/abs/2018MNRAS.476.2092L} {476, 2092}

\bibitem[\protect\citeauthoryear{{Lowing}, {Jenkins}, {Eke}  \&
  {Frenk}}{{Lowing} et~al.}{2011}]{lowing2011}
{Lowing} B.,  {Jenkins} A.,  {Eke} V.,   {Frenk} C.,  2011, \mn@doi [\mnras]
  {10.1111/j.1365-2966.2011.19222.x}, \href
  {http://adsabs.harvard.edu/abs/2011MNRAS.416.2697L} {416, 2697}

\bibitem[\protect\citeauthoryear{Mullen}{Mullen}{1966}]{Mullen1966}
Mullen J.~A.,  1966, SIAM Journal on Applied Mathematics, 14, 1152

\bibitem[\protect\citeauthoryear{{Navarro}, {Frenk}  \& {White}}{{Navarro}
  et~al.}{1997}]{NFW1997}
{Navarro} J.~F.,  {Frenk} C.~S.,   {White} S.~D.~M.,  1997, \mn@doi [\apj]
  {10.1086/304888}, \href {http://adsabs.harvard.edu/abs/1997ApJ...490..493N}
  {490, 493}

\bibitem[\protect\citeauthoryear{{Ngan}, {Bozek}, {Carlberg}, {Wyse}, {Szalay}
  \& {Madau}}{{Ngan} et~al.}{2015}]{Ngan2015}
{Ngan} W.,  {Bozek} B.,  {Carlberg} R.~G.,  {Wyse} R.~F.~G.,  {Szalay} A.~S.,
  {Madau} P.,  2015, \mn@doi [\apj] {10.1088/0004-637X/803/2/75}, \href
  {http://adsabs.harvard.edu/abs/2015ApJ...803...75N} {803, 75}

\bibitem[\protect\citeauthoryear{{Olver}, {Olde Daalhuis}, {Lozier},
  {Schneider}, {Boisvert}, {Clark}, {Miller}  \& {Saunders}}{{Olver}
  et~al.}{2016}]{dlmf}
{Olver} F.~W.~J.,  {Olde Daalhuis} A.~B.,  {Lozier} D.~W.,  {Schneider} B.~I.,
  {Boisvert} R.~F.,  {Clark} C.~W.,  {Miller} B.~R.,   {Saunders} B.~V.,  2016,
  {NIST Digital Library of Mathematical Functions}, \url
  {http://dlmf.nist.gov/}

\bibitem[\protect\citeauthoryear{{Plummer}}{{Plummer}}{1911}]{Plummer1911}
{Plummer} H.~C.,  1911, \mn@doi [\mnras] {10.1093/mnras/71.5.460}, \href
  {http://adsabs.harvard.edu/abs/1911MNRAS..71..460P} {71, 460}

\bibitem[\protect\citeauthoryear{{Rahmati} \& {Jalali}}{{Rahmati} \&
  {Jalali}}{2009}]{rahmati2009}
{Rahmati} A.,  {Jalali} M.~A.,  2009, \mn@doi [\mnras]
  {10.1111/j.1365-2966.2008.14226.x}, \href
  {http://adsabs.harvard.edu/abs/2009MNRAS.393.1459R} {393, 1459}

\bibitem[\protect\citeauthoryear{Saad \& Hall}{Saad \& Hall}{2003}]{Saad2003}
Saad N.,  Hall R.~L.,  2003, Journal of Physics A: Mathematical and General,
  36, 7771

\bibitem[\protect\citeauthoryear{{Saha}}{{Saha}}{1993}]{Saha1993}
{Saha} P.,  1993, \mn@doi [\mnras] {10.1093/mnras/262.4.1062}, \href
  {http://adsabs.harvard.edu/abs/1993MNRAS.262.1062S} {262, 1062}

\bibitem[\protect\citeauthoryear{{Tremaine}, {Richstone}, {Byun}, {Dressler},
  {Faber}, {Grillmair}, {Kormendy}  \& {Lauer}}{{Tremaine}
  et~al.}{1994}]{Tremaine1994}
{Tremaine} S.,  {Richstone} D.~O.,  {Byun} Y.-I.,  {Dressler} A.,  {Faber}
  S.~M.,  {Grillmair} C.,  {Kormendy} J.,   {Lauer} T.~R.,  1994, \mn@doi [\aj]
  {10.1086/116883}, \href {http://adsabs.harvard.edu/abs/1994AJ....107..634T}
  {107, 634}

\bibitem[\protect\citeauthoryear{{Veltmann}}{{Veltmann}}{1979}]{Ve79}
{Veltmann} U.~I.~K.,  1979, \sovast, \href
  {http://adsabs.harvard.edu/abs/1979SvA....23..551V} {23, 551}

\bibitem[\protect\citeauthoryear{{Weinberg}}{{Weinberg}}{1999}]{Wein99}
{Weinberg} M.~D.,  1999, \mn@doi [\aj] {10.1086/300669}, \href
  {http://adsabs.harvard.edu/abs/1999AJ....117..629W} {117, 629}

\bibitem[\protect\citeauthoryear{{Zhao}}{{Zhao}}{1996}]{Zhao1996}
{Zhao} H.,  1996, \mn@doi [\mnras] {10.1093/mnras/278.2.488}, \href
  {http://adsabs.harvard.edu/abs/1996MNRAS.278..488Z} {278, 488}

\bibitem[\protect\citeauthoryear{{de Zeeuw}}{{de Zeeuw}}{1985}]{dezeeuw1985}
{de Zeeuw} T.,  1985, \mn@doi [\mnras] {10.1093/mnras/216.2.273}, \href
  {http://adsabs.harvard.edu/abs/1985MNRAS.216..273D} {216, 273}

\makeatother
\end{thebibliography}

\onecolumn
\appendix

\section{Derivation of general expressions}\label{sec::general_derivation}
As indicated in \eqref{eqn::gk_with_K}, the functions $g_n(k)$ are a
weighted sum of the functions $\mathcal{K}_j(k)$ \eqref{eqn::curly_K}. Writing
$\mathcal{K}_j(k)$ using an integral representation of the modified Bessel
function $K_\nu(k)$ \citep[10.32.10]{dlmf}, and writing the weights $c_{nj}$ using the
polynomial \eqref{eqn::jacobi}, we have the following integral
expression for the functions $g_n(k)$,
\begin{equation}
g_n(k) = \frac{\mu \: k^{\mu+2\nu-1}}{2^{\mu+2\nu-1} \: \Gamma(\mu+\nu)} \: \int_0^\infty \dif t \: t^{-\nu-1} \: \expe^{-t-\frac{k^2}{4t}} \: f_n(t),
\end{equation}
where
\begin{equation}\label{eqn::fn_def}
f_n(t) = {}_2 \mathcal{F}_2\left(\left.%
      \begin{matrix}
        -n, n+2\mu+2\nu-1 \\
        \mu,\mu+\nu
      \end{matrix}%
    \right| t \right),
\end{equation}
which is essentially the Jacobi polynomial \eqref{eqn::jacobi}
together with an additional factor multiplying each term that arises
from the inner product calculation \eqref{eqn::Koverlap}.  We can now
insert these expressions for $g_n(k)$ into \eqref{eqn::familyA}, using
\citet[6.631(1)]{gradshteyn4} to evaluate the integral over the Bessel
$J$-function, to obtain integral expressions for ${\Phi}_{nl}$ and
${\rho}_{nl}$,
\begin{equation}\label{eqn::phi_int_form}
{\Phi}_{nl} = \frac{r^l}{\Gamma(\mu)} \: \int_0^\infty \dif t \: t^{\mu-1} \: \expe^{-t} \: f_n(t) \:
  {}_1 \mathcal{F}_1\left(\left.%
      \begin{matrix}
        \mu+\nu \\
        \mu+1
      \end{matrix}%
    \right| -z^2 t \right),
\end{equation}
\begin{equation}\label{eqn::rho_int_form}
{\rho}_{nl} = \frac{r^{l-2+1/\alpha}}{\Gamma(\mu+1)} \: \int_0^\infty \dif t \: t^{\mu} \: \expe^{-t} \: f_n(t) \:
  {}_1 \mathcal{F}_1\left(\left.%
      \begin{matrix}
        \mu+\nu+1 \\
        \mu+1
      \end{matrix}%
    \right| -z^2 t \right).
\end{equation}
%
%
As these expressions are in integral form, they are not yet
useful\footnote{Although it is interesting to note that a valid --
  though not necessarily biorthogonal -- potential-density pair would be
  given by replacing the integrand (apart from the confluent
  hypergeometric function) by any function of $t$.}. We proceed with a
method based on generating functions. By substituting the appropriate
values into \citet[Eq.~26]{Chaundy1943}, we can find a generating
function for $f_n(t)$, noting that the result fortuitously simplifies
from a ${}_2\mathcal{F}_2$ to a ${}_1\mathcal{F}_1$ function,
\begin{equation}\label{eqn::fn_gen_fun}
  \sum_{n=0}^\infty \frac{(2\mu+2\nu-1)_n}{n!} f_n(t) x^n = (1-x)^{1-2\mu-2\nu}%
    {}_1 \mathcal{F}_1\left(\left.%
      \begin{matrix}
        \mu+\nu-1/2 \\
        \mu
      \end{matrix}%
    \right| \frac{-4tx}{(1-x)^2} \right).
\end{equation}
This expression can be used in \eqref{eqn::phi_int_form}, and the
resulting integral over the pair of ${}_1\mathcal{F}_1$ functions can
be evaluated using \citet[Eq.~2.2]{Saad2003}, to give
\begin{multline}\label{eqn::G_nu_mu}
\sum_{n=0}^\infty \frac{(2\mu+2\nu-1)_n}{n!} {\Phi}_{nl} x^n = \frac{r^l}{(1-x)^{2\mu+2\nu-1}}%
    F_2\left(\left.%
      \begin{matrix}
        \mu;\mu+\nu-1/2,\mu+\nu \\
        \mu,\mu+1
      \end{matrix}%
    \right| \frac{-4x}{(1-x)^2}, -z^2 \right) \\
  = \frac{r^l}{(1+x)^{2\mu+2\nu-1}}%
  F_1\left(\left.%
      \begin{matrix}
        \mu+\nu;\mu+\nu-1/2,1/2-\nu \\
        \mu+1
      \end{matrix}%
    \right| -\left(\frac{1-x}{1+x}\right)^2z^2, -z^2 \right)
\end{multline}
where $F_1$ and $F_2$ are Appell hypergeometric functions, and the
$F_2 \to F_1$ reduction \citep[16.16.3]{dlmf} is justified because the first and fourth
arguments of the $F_2$ are equal.
An $F_1(a;b_1,b_2;c;z)$ function simplifies to a ${}_2\mathcal{F}_1$
function \citep[16.16.1]{dlmf} if $b_1+b_2=c$, and we note that second
parameter of the $F_1$ in \eqref{eqn::G_nu_mu} would need to be
increased by $1$ in order to satisfy this condition. To accomplish this
we make use of the following relation, derivable from the $F_1$
contiguous relations \citep{Mullen1966},
\begin{equation}
  F_1\left(\left.%
      \begin{matrix}
        a;b_1+1,b_2 \\
        c
      \end{matrix}%
    \right| s, t \right)
  =
  F_1\left(\left.%
      \begin{matrix}
        a;b_1,b_2 \\
        c
      \end{matrix}%
    \right| s, t \right)
  +
  \frac{s}{b_1} \pd{}{s}
    F_1\left(\left.%
      \begin{matrix}
        a;b_1,b_2 \\
        c
      \end{matrix}%
    \right| s, t \right).
\end{equation}
Applying this relation to \eqref{eqn::G_nu_mu}, simplifying both sides
of the equation, and applying the now-valid
$F_1 \to {}_2\mathcal{F}_1$-reduction formula, we obtain
\begin{equation}
  \sum_{n=0}^\infty \frac{(2\mu+2\nu)_n}{n!} \left({\Phi}_{nl} - {\Phi}_{n+1,l}\right) x^n = 2 \: r^l \: (1+x)^{-2\mu-2\nu} \: (1+z^2)^{-\mu-\nu} \: %
  {}_2 \mathcal{F}_1\left(\left.%
      \begin{matrix}
        \mu+\nu,\mu+\nu+1/2 \\
        \mu+1
      \end{matrix}%
    \right| \frac{4x\chi}{(1+x)^2} \right).
\end{equation}
This generating function is also a special case of
\citet[Eq.~26]{Chaundy1943} and in fact turns out to be a generating
function for the Jacobi polynomials, so we finally obtain
\begin{equation}
{\Phi}_{nl} - {\Phi}_{n+1,l} = \frac{2 \: n!}{(\mu+1)_n} \: \frac{r^l}{(1+z^2)^{\mu+\nu}} \: P_n^{(\mu+2\nu-1,\mu)}(\xi).
\end{equation}
A similar method can be used for ${\rho}_{nl}$, starting from
\eqref{eqn::rho_int_form} and applying the generating function
\eqref{eqn::fn_gen_fun}, then integrating using
\citet[Eq.~2.2]{Saad2003} and applying the $F_2\to F_1$
transformation, to give
\begin{equation}
\sum_{n=0}^\infty \frac{(2\mu+2\nu-1)_n}{n!} {\rho}_{nl} x^n = \frac{r^{l-2+1/\alpha}}{(1+z^2)^{\mu+\nu+1}(1-x)^{2\mu+2\nu-1}}%
  F_1\left(\left.%
      \begin{matrix}
        \mu+\nu-1/2;-\nu,\mu+\nu+1 \\
        \mu
      \end{matrix}%
    \right| \frac{-4x}{(1-x)^2}, \frac{-4x}{(1-x)^2(1+z^2)} \right).
\end{equation}
This time we note that the fourth parameter of the $F_1$ needs to be
increased by $1$ in order to reduce it to an ${}_2\mathcal{F}_1$. To
accomplish this, we note the following $F_1$ contiguous relation \citep{Mullen1966},
\begin{equation}
  F_1\left(\left.%
      \begin{matrix}
        a;b_1,b_2 \\
        c
      \end{matrix}%
    \right| s, t \right)
  = \frac{c-a}{c}%
  F_1\left(\left.%
      \begin{matrix}
        a;b_1,b_2 \\
        c+1
      \end{matrix}%
    \right| s, t \right)
  +\frac{a}{c}%
    F_1\left(\left.%
      \begin{matrix}
        a+1;b_1,b_2 \\
        c+1
      \end{matrix}%
    \right| s, t \right).
\end{equation}
Having applied this, we can use the $F_1\to{}_2\mathcal{F}_1$
transformation twice, giving
\begin{multline}
\sum_{n=0}^\infty \frac{(2\mu+2\nu-1)_n}{n!} {\rho}_{nl} x^n = \frac{r^{l-2+1/\alpha}}{(1+z^2)^{\mu+\nu+1}}%
\left[\frac{1/2-\nu}{\mu}(1+x)^{1-2\mu-2\nu}%
  {}_2 \mathcal{F}_1\left(\left.%
      \begin{matrix}
        \mu+\nu-1/2,\mu+\nu+1 \\
        \mu+1
      \end{matrix}%
    \right| \frac{4x\chi}{(1+x)^2} \right)\right.\\
\left.+ \frac{\mu+\nu-1/2}{\mu} (1-x)^2 (1+x)^{-1-2\mu-2\nu}%
  {}_2 \mathcal{F}_1\left(\left.%
      \begin{matrix}
        \mu+\nu+1/2,\mu+\nu+1 \\
        \mu+1
      \end{matrix}%
    \right| \frac{4x\chi}{(1+x)^2} \right)
\right].
\end{multline}
We apply \citet[15.5.15]{dlmf} to the first ${}_2\mathcal{F}_1$, which
turns it into two \citet{Chaundy1943}-style generating functions for
the Jacobi polynomials $P_n^{(\mu+2\nu-1,\mu-1)}(\xi)$ and
$P_n^{(\mu+2\nu-2,\mu)}(\xi)$; the second ${}_2\mathcal{F}_1$ is a
\citet{Chaundy1943}-style generating function multiplied by a factor
of $(1-x)^2$ and so produces terms proportional to
$P_n^{(\mu+2\nu,\mu)}(\xi)$, $P_{n-1}^{(\mu+2\nu,\mu)}(\xi)$ and
$P_{n-2}^{(\mu+2\nu,\mu)}(\xi)$; hence we obtain a sum of five Jacobi
polynomials with various parameters. We must then apply
\citet[18.9.3,18.9.5]{dlmf} several times to simplify the expression
to give the final result, namely
\begin{multline}
  {\rho}_{nl} = \frac{n!(n+\mu)}{\mu(\mu+\nu)(2n+2\mu+2\nu-1)(\mu+1)_n} \:\: \frac{r^{l-2+1/\alpha}}{(1+z^2)^{\mu+\nu+1}}%
  \left[(n+2\mu+2\nu-1)(n+\mu+\nu)P_n^{(\mu+2\nu-1,\mu)}(\xi)\right.\\
  - \left.(n+\mu+2\nu-1)(n+\mu+\nu-1)P_{n-1}^{(\mu+2\nu-1,\mu)}(\xi)\right].
\end{multline}
For simplicity, the expressions \eqref{eqn::phi_hat} and
\eqref{eqn::rho_hat} in the main body of the paper are written using a
different normalisation.

\section{Limiting forms}\label{sec::limiting}

In certain cases the density $\rho_{nl}$ and associated constants
$N_{nl}$, $K_{nl}$ must be modified, as they diverge or become
zero. Modification is required when two conditions are satisfied:
$n=l=0$, and $\alpha + \nu = 1/2$. Because of the pre-existing
constraints on $\nu$ and $\alpha$, this means that the only cases
affected are $1/2 \leq \alpha < 1$ and $-1/2 < \nu \leq 0$ (this
includes the basis set with zeroth order the modified Hubble
profile). We set $n=l=0$ first, then evaluate the following limits as
$\nu \to 1/2 - \alpha$, making use of $\lim_{x\to 0}
\left[x\Gamma(x)\right] = 1$,
\begin{equation}
\begin{split}
\lim_{\nu\to 1/2-\alpha} \left[K_{00}\rho_{00}\right] & = -\frac{1}{8\pi\alpha} \: \frac{r^{-2+1/\alpha}}{(1+z^2)^{3/2}}, \\
\lim_{\nu\to 1/2-\alpha} \left[K_{00}N_{00}\right] & = -\frac{\alpha}{4} \: \cosec{(\pi\alpha)}.
\end{split}
\end{equation}
For these special cases the orthogonality relation
\eqref{eqn::orthogonal} must be multiplied through by $K_{00}$ in
order to have meaning. Note that the result depends on the order in
which the limits $n,l\to 0$ and $\nu \to 1/2-\alpha$ were taken, so
the same order must be used for both quantities, otherwise
\eqref{eqn::orthogonal} will not hold.

\bsp	
\label{lastpage}
\end{document}